\documentclass[aps,pre,twocolumn,showpacs,groupedaddress]{revtex4-1}

\usepackage{amssymb,amsmath,graphicx,bm,textcomp,xcolor,commath,bm}
\usepackage{ulem}
\usepackage[utf8]{inputenc}

\def \bff {{\bf f}}
\def \bft {{\bf t}}
\def \bftbar {{\bf {\overline t}}}
\def \bfnbar {{\bf {\overline n}}}
\def \bfn {{\bf n}}
\def \bfN {{\bf N}}
\def \bfT {{\bf T}}
\def \bfe {{\bf e}}
\def \bfp {{\bf p}}
\def \bfr {{\bf r}}

\newcommand{\red}[1]{\textcolor{black}{#1}}

\begin{document}

\title{Metastability of a periodic network of threads:\\ what are the shapes of a knitted fabric ?}

\author{J\'er\^ome Crassous} \affiliation{Univ Rennes, CNRS, IPR (Institut de Physique de Rennes) - UMR 6251, F-35000 Rennes, France}\email{jerome.crassous@univ-rennes.fr}

\author{Samuel Poincloux} \affiliation{Department of Physical Sciences, Aoyama Gakuin University, 5-10-1 Fuchinobe, Sagamihara, Kanagawa 252-5258, Japan}

\author{Audrey Steinberger} \affiliation{Univ Lyon, ENS de Lyon, CNRS, Laboratoire de Physique, F-69342 Lyon, France}

\date{\today}

\begin{abstract}
Knitted fabrics are metamaterials with remarkable mechanical properties, such as extreme deformability and multiple history-dependent rest shapes. This letter shows that those properties may stem from a continuous set of metastable states for a fabric free of external forces. This is evidenced through experiments, numerical simulations and analytical developments. Those states arise from the frictional contact forces acting in the braid zone where the threads interlace and follow a line in the configuration space accurately described by a 2D-elastica model. The friction coefficient sets a terminal point along this line, and the continuous set of metastable states is obtained by varying the braid inclination while contact forces remain on the friction cone. 

\end{abstract}
\maketitle


Assemblies of long, flexible, and intertwined fibers with frictional contacts are materials involved in various phenomena, including surgical or shoe knots \cite{johanns_strength_2023,daily_roles_2017}, nests and self-assembled natural structures \cite{weiner_mechanics_2020,verhille.2017}, nonwoven fabrics with a wealth of applications \cite{albrecht_nonwoven_2006}, or the degradation of ancient manuscripts \cite{vibert_relationship_2024}.
Despite being essential for most applications, the mechanical response of fiber assemblies is intrinsically non-linear, dissipative, and history-dependent, stemming from the fibers' slenderness and the frictional contacts. 
Providing quantitative mechanical predictions for the assembly from the properties of the fibers remains a theoretical and numerical challenge, with recent advances made in simplified situations with tight geometries \cite{poincloux.2021,seguin_twist_2022,chopin.2024}.
One particular class of ordered fiber assemblies, textiles, have tremendous industrial importance in manufacturing \cite{long_design_2005} or geo-engineering \cite{koerner_geotextiles_2016}. 
They also recently gained a renewed interest as metamaterials with extensive programmability \cite{poincloux.2018b,singal_programming_2024} for emerging soft robotics and smart textile applications \cite{chen_smart_2020,sanchez_textile_2021}.
However, the prediction of basic properties, like the rest shape of a knitted fabric given the length by stitch of its constitutive yarn, is an old but still open question \cite{munden_geometry_1959,lanarolle_geometry_2021} even though a reproducible state can be achieved after repeated multidirectional stretching \cite{allan_heap_prediction_1983}. 
One possible way of progress may emerge from yarn-level simulation of knitted fabrics for which the computer graphics community made enormous progress \cite{kaldor_simulating_2008,sperl_estimation_2022,ding_unravelling_2023}, but the dynamics usually rely on viscous dissipative forces at the contacts, ill-adapted to capture rest shapes arrested by dry friction. 
Nonetheless, recent numerical advances allow the combination of large fiber displacements with frictional interactions \cite{cirio_yarn-level_2017,li_implicit_2018,liu_computational_2018,crassous_discrete_2023} and open the way to explore quantitatively the mechanics and stability of frictional fiber assemblies \cite{sano_randomly_2023}. The complex geometry of the contact zones between fibers makes exact theoretical modeling extremely complicated.
In this letter, we show that a simplified description of these zones can faithfully reproduce the rest shapes of a knitted fabric. The postulate of a single form of equilibrium must be abandoned. Even with zero external stresses ($\sigma^{\mathrm{ext}} = 0$), the solid friction between the threads stabilizes the materials in various metastable states depending on the system's history.


\begin{figure}[t]
\centering
\includegraphics[width=\columnwidth]{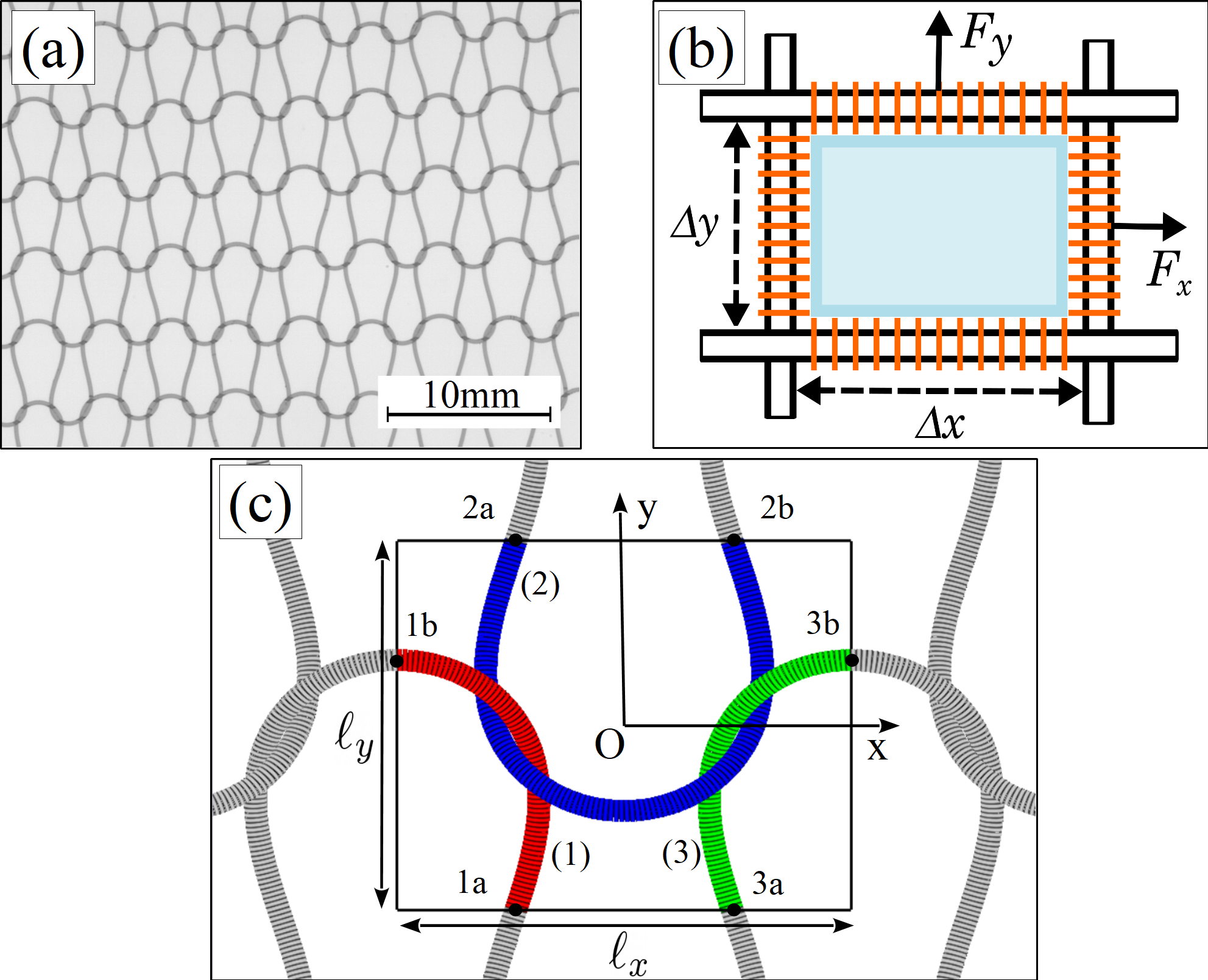}
\caption{(a) Photo of a Jersey stitch knit.  (b) Experimental set-up. The knitted fabric (light blue) is attached with metallic staples (orange) that can freely slide over $4$ cylindrical bars linked to a biaxial tensile machine. 
(c) Geometry and numerical model. A rectangular cell of size $\ell_x\times\ell_y$ is composed of 3 threads $(1),(2)\,\&\,(3)$. Segments of cylinders composing the threads are colored. Dots $1a$, $1b$,... are the endpoints of the different threads, which are constrained to be on the cell boundaries with periodic boundary conditions.}
\label{fig:exp}
\end{figure}

In this study, we use a Jersey stitch knit, which is both simple and widely used. It consists of a single yarn forming interlocked loops as in Fig.~\ref{fig:exp}(a). Experimentally, we make a knitted fabric of $70 \times 70$ stitches from a polyamide (Nylon) thread (Madeira Monofil n°40, $E=1.05\,\mathrm{GPa}$, $\mu=0.5$) of diameter $d=0.155~\,\mathrm{mm}$. The length of thread per stitch is $\ell=9.7\,\mathrm{mm}$. The central $N \times N$ stitches ($N=50$) are attached to a biaxial tensile machine (Fig.~\ref{fig:exp}.b), where the spacing $\Delta x$ along the courses and $\Delta y$ along the wales can be varied by stepper motors. The forces per row $f_x := F_x/N$ and columns $f_y := F_y/N$ are measured using strain gauge force sensors. The network's periodicity $(\ell_x,\ell_y)$ is obtained from images recorded with a camera. The network is also studied using numerical simulations based on Discrete Elastic Rods (DER) coupled with dry contacts with a friction coefficient $\mu$~\cite{crassous_discrete_2023,SM}. Threads are decomposed into segments of circular cylinders connected by springs, which account for the elastic forces of traction, flexion, and torsion. A mesh comprises 3 rods as shown in fig~\ref{fig:exp}.c. The endpoints of these 3 wires are constrained to the $\pm \ell_x/2$ or $\pm \ell_y/2$ planes. Periodic boundary conditions in terms of positions and forces are applied at the junctions between the strands: for example, the strands $1$ and $2$ satisfy $\bfr_{2a}-\bfr_{1a}= \ell_y \bfe_y$, $\alpha_{2a}=\alpha_{1a}$, $\bfr^{'}_{2a}=\bfr^{'}_{1a}$ and $\bfr^{''}_{2a}=\bfr^{''}_{12}$. $\alpha$ is the torsion angle, $^{'}$ and $^{''}$ are the 1st and 2nd derivatives with respect to the curvilinear abscissa $s$. Conditions on the derivatives ensure the continuity of the forces and moments. 

\begin{figure}[t]
\centering
\includegraphics[width=\columnwidth]{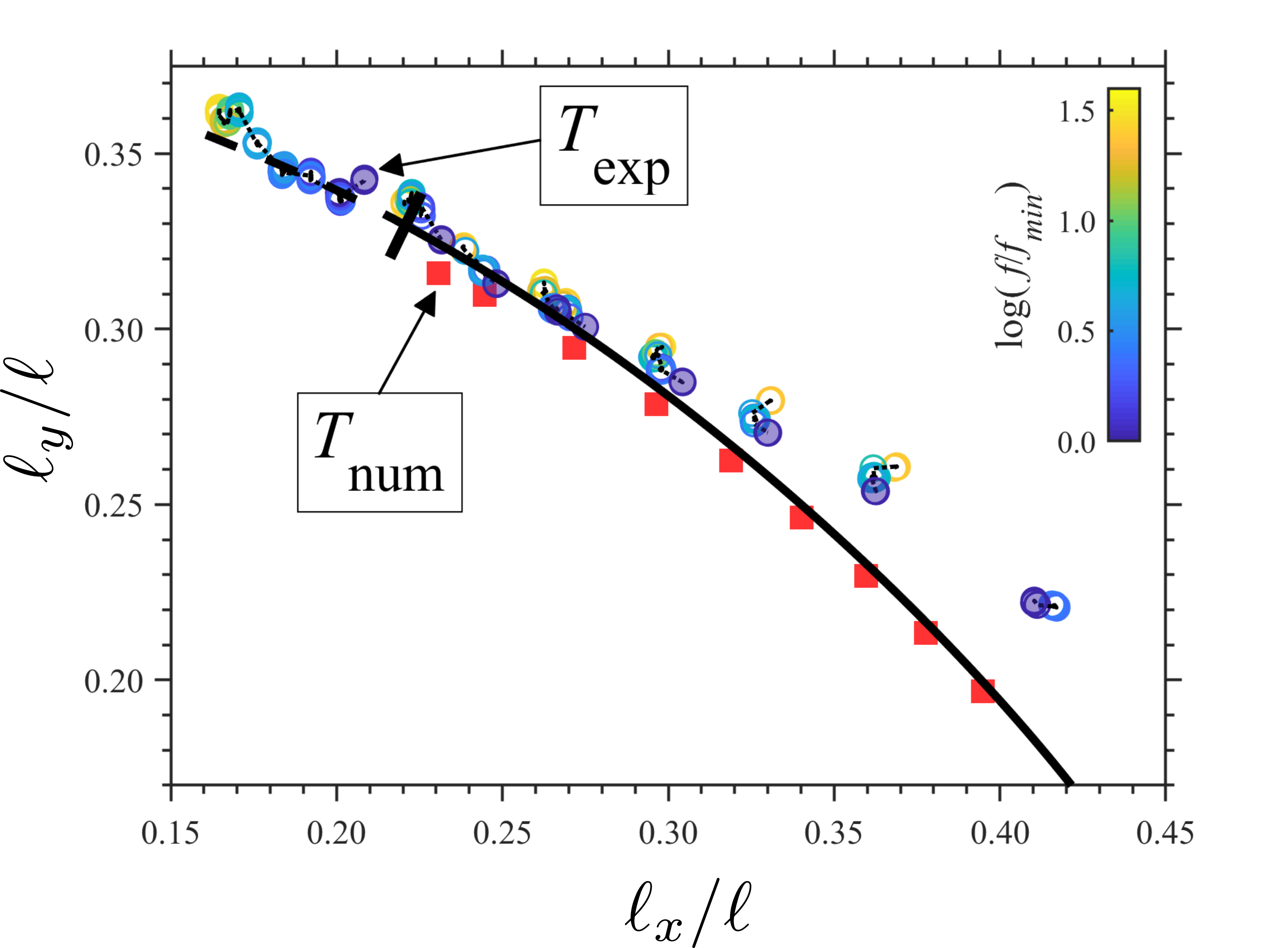}
\caption{Relaxed configurations of a knitted fabrics ($\ell/d=62.5$, $\mu=0.5$). (open circles): Experimental configurations during relaxation, color-coded by $\log(f/f_{min})$. The dotted lines are guidelines. (filled disks): Relaxed, experimental configurations; (red squares): Configurations obtained from DER simulations. There are no relaxed configurations above or to the left of the terminal points $T$, noted $T_{\mathrm{exp}}$ (resp. $T_{\mathrm{num}}$) for experimental (resp. numerical) data. (black line): Possible configurations expected from a 2D-elastica model with no applied stress, with the predicted terminal point separating \red{the accessible (plain) and forbidden (dashed) regions}.}
\label{fig:compa}
\end{figure}

We identify the rest states at $\sigma^{\mathrm{ext}}=0$ as follows. Experimentally, the knitted fabric is stretched to an initial state $(\Delta x_0,\Delta y_0)$, then $\Delta x$ and $\Delta y$ are varied to reduce $f:=(f_x^2+f_y^2)^{1/2}$ until it becomes lower than $f_{min}=8$~mN. Numerically, starting from a given stretched configuration, the stitch length is varied by $\delta \ell_i=-\lambda f_i$, with $i=x,y$ and $\lambda$ a numerical constant, until the force is smaller to a given threshold. Fig.\ref{fig:compa} shows the mesh sizes obtained following this experimental and numerical protocol.
Firstly, the rest shape of the knitted fabric is not uniquely defined and is strongly dependent on the initial state. Secondly, these shapes belongs to a curve which is an attractor in the space of $(\ell_x,\ell_y)$ configurations. This curve ends at a terminal point $T:=(\ell_{x_T},\ell_{y_T})$. A knit verifying $\ell_y>\ell_{y_T}$ or $\ell_x<\ell_{x_T}$ is impossible without external forces. Finally, the experimental and numerical data agreement is very satisfactory: modeling a polyamide thread with an elastic fiber without cross-sectional deformations is a reasonable hypothesis. Plastic deformations occurring during knitting, leading to stress-free yarn no longer ideally rectilinear, underlie the observed differences.  

\begin{figure}[t]
\centering
\includegraphics[width=\columnwidth]{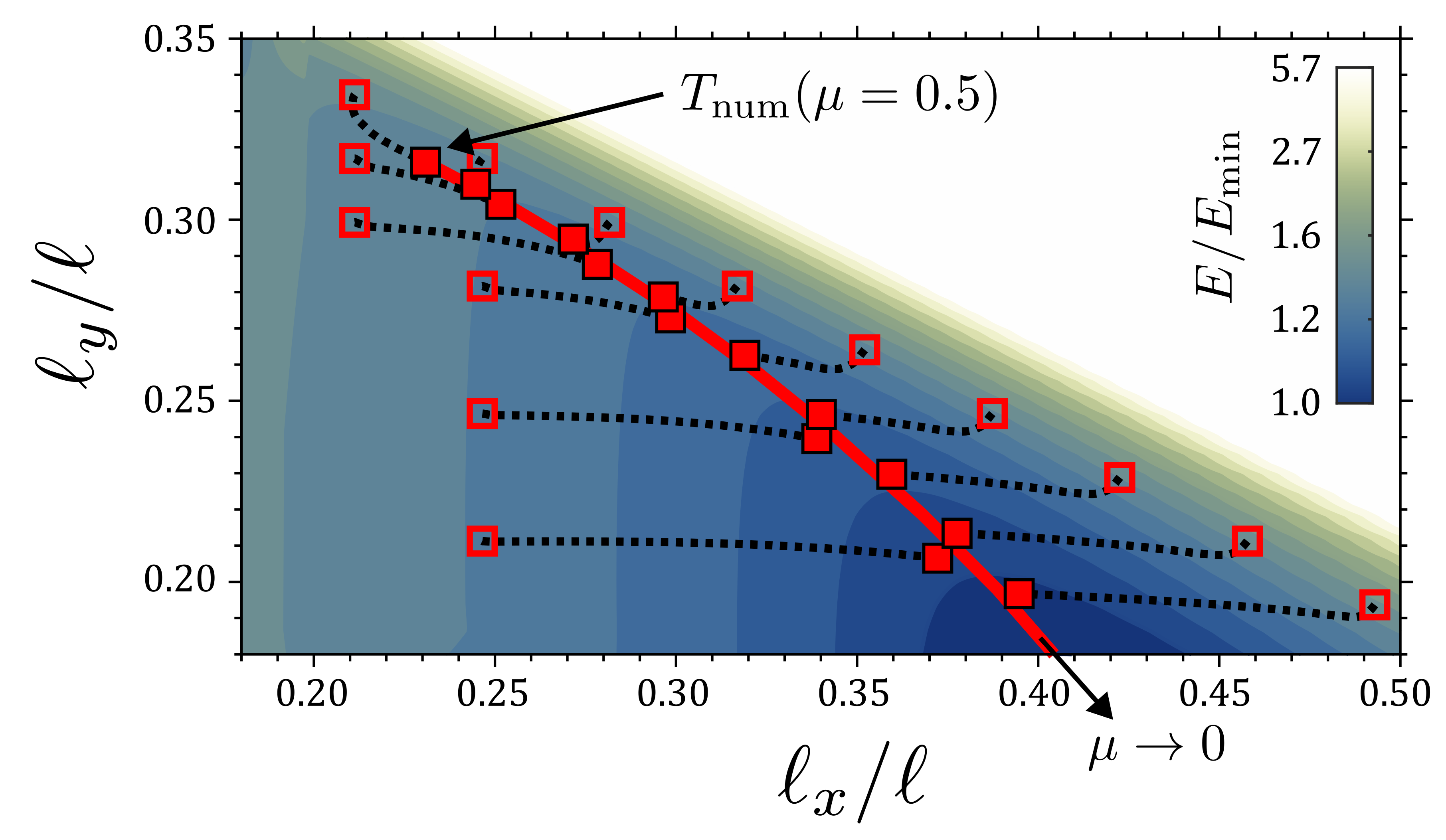}
\caption{Elastic energy $E$ (color bar) for cells \red{of imposed size} made of frictionless threads ($\mu=0$, $\ell/d=62.5$). \red{The energy is normalized by $E_{\mathrm{min}}$, the minimum energy reached by a frictionless stitch without external stress.} (Dotted lines): Relaxations of fabric with frictional thread ($\mu=0.5$) between an initial configuration (open red square) and a relaxed configuration (filled red square). $T_{\mathrm{num}}(\mu=0.5)$ is the terminal point. (Plain red curve): Relaxation of a fabric, initially at $T_{\mathrm{num}}(\mu=0.5)$, when $\mu$ is gradually decreased.}
\label{fig:energy}
\end{figure}

To describe the set of rest states, we consider the elastic energy $E(\ell_x,\ell_y)=E^{(b)}+E^{(t)}+E^{(s)}$, with respectively the bending, twisting and stretching energy. It is obtained from DER simulations for a unit cell of frictionless fabric (Fig.~\ref{fig:energy}).  A valley that descends towards small $\ell_y$ and large $\ell_x$ separates the traction zones (large $\ell_x$ or $\ell_y$) from the compressive zones (small $\ell_x$ or $\ell_y$). The knitted fabric relaxes its bending energy by aligning its yarns parallel to $\bfe_x$. The milder slope in this energy landscape acts as an attractor line for frictional knitted fabrics with $\mu\neq 0$. Stretched or compressed fabrics relax (Fig.~\ref{fig:energy}, dotted lines) and stop close to the valley floor. Due to the non-conservative nature of frictional forces, the trajectories obtained at $\mu\neq 0$ are not maximal slopes of $E(\ell_x,\ell_y)$. 

We then obtain the rest configurations and terminal points $T_{\mathrm{num}}$ as functions of $\mu$ (Fig.~\ref{fig:energy}, continuous red line). The endpoints $T_{\mathrm{num}}(\mu)$ sit along a line on the valley floor and go down the milder slope as $\mu$ decreases. This line is the attractor for the relaxing trajectories which stop in its vicinity. As $\mu\to0$, this line ends because contacts between non-successive rows occur, but this limit is not discussed here.  


Therefore, describing the set of metastable configurations requires understanding i) the shape of the friction-independent valley; ii) how the friction $\mu$ controls the position of the terminal point $T$ on the line of milder slope.

\begin{figure}[t]
\centering
\includegraphics[width=\columnwidth]{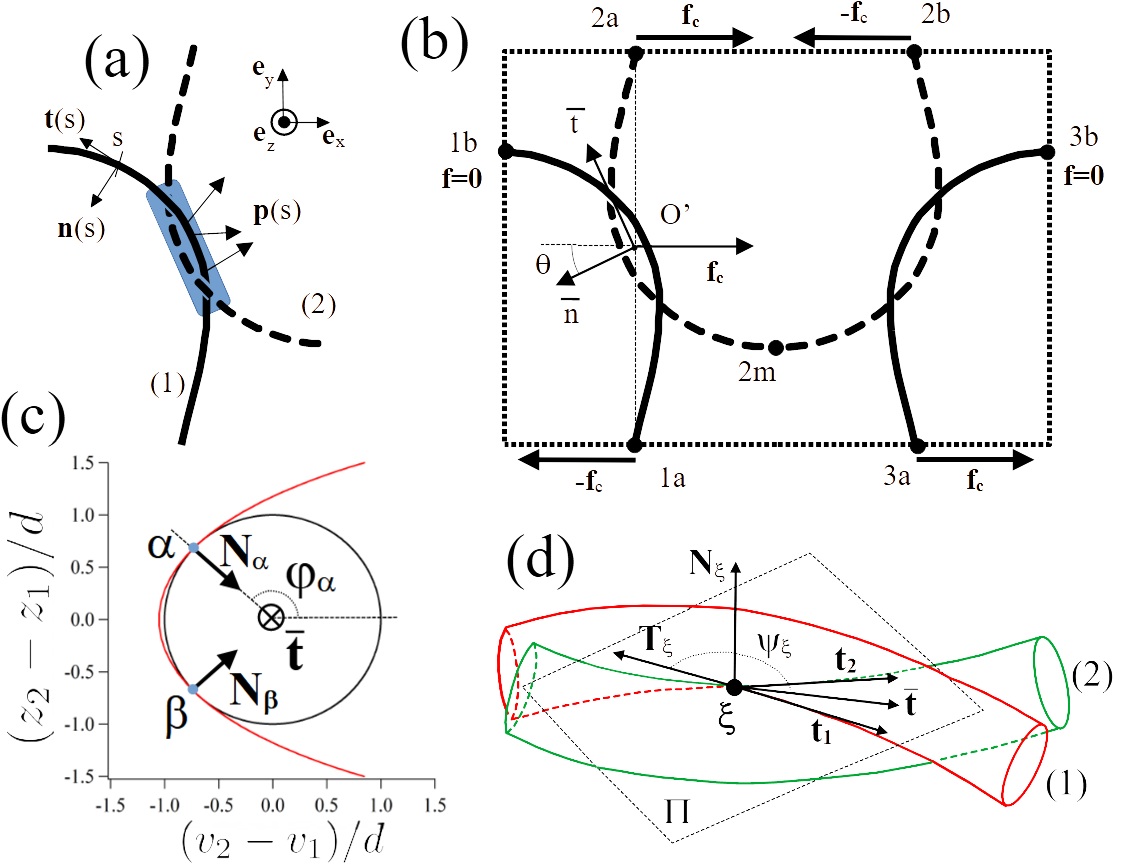}
\caption{(a) Contact forces $\bfp(s)$ acting in the braid zone (blue rectangle). (b) Forces acting on one mesh for a rest state ${\bf \sigma^{\mathrm{ext}}}=0$. The total forces on each side is $0$, and the internal forces acting on threads are $0$ or $\pm \bf{f_c}$. The points $(1a)$ and $(2a)$ are on the vertical line of middle point $O'$, and $\bftbar$ is the braid direction. (c) (Red line): Projection of the vector $\bfr_2-\bfr_1$, joining the centerlines, on the planes perpendicular to $\bftbar$. $v$ is the coordinate along an axis  of direction $-\bfnbar := \bftbar \times \bfe_z$. The circle is of unit radius.  Contact occurs in two points $\alpha$ and $\beta$ (blue dots) that makes angles $\varphi_\alpha$ and $\varphi_\beta=-\varphi_\alpha$ with the $z=0$ plane. (d) Orientations of the contact force at a contact point $\xi$. The plane $\Pi$ containing $\bft_1$ and $\bft_2$ is normal to $\bfN_\xi$. $\bfT_\xi$ is the direction of the tangential contact force, which makes an angle $\Psi_\xi$ with $\bftbar$.} 
\label{fig:cross}
\end{figure}

Interactions between the yarns take place in the braid 
where they become entangled, creating both normal and tangential forces to the threads. To describe this zone, schematically depicted in Fig.~\ref{fig:cross}.a, we introduce the curvilinear abscissa $s$ along the centerline of thread (1), and  $\bfp(s)$ the linear density of contact force exerted by (2) on (1). Because friction is present, $\bfp(s)$ is not necessarily aligned to the normal vector $\bfn(s)$. We define 
\begin{equation}
    \bff_c=\int_{s_{in}}^{s_{out}} \bfp(s) ds\label{eq:fc}
\end{equation}
the resultant of contact forces with $s_{in}$ the entrance of the contact zone: $\bfp(s_{in})\ne 0$ and $\bfp(s)=0$  for $s<s_{in}$, with reciprocal definition for $s_{out}$. For any rest state ${\bf \sigma^{\mathrm{ext}}}=0$, the sum of total forces acting on each side of the unit cell must be $0$ (see Fig.\ref{fig:cross}.b). Along the row, the external forces acting on points $(1b)$ and $(3b)$ vanish $\bff_{1b}=\bff_{3b}=0$, while along the column $\bff_{2a}+\bff_{2b} = \bff_{1a}+\bff_{3a}=0$. Periodicity and symmetry along the $y$ axis imply, $\bff_{1a}=-\bff_{2a}$ and $\bff_{1a}=f_{1a}\bfe_x$. Finally, equilibrium of thread $(1)$ with $\bff_{1a}+\bff_{1b}+\bff_{c}=0$ results in an horizontal contact force $\bff_c\cdot \bfe_y=0$ and $\bff_{1a}=-\bff_{c}$. The results of the DER simulations fulfilled all those requirements.

Let $s_c$ be the arbitrary abscissa of contact at which $\bff_c$ is applied. In the limit $d/\ell\ll 1$, we consider the bi-dimensional problem of finding the value of $f_c:=\Vert \bff_c \Vert$ such that the 2D elastic curve $\bfr(s)$ representing the strand (1) is at equilibrium. The curve must satisfy the applied external force $\bff_c$ at $s=s_c$ and $-\bff_c$ at $s=s_{1a}=0$, and verifying $dy/ds=0$ at $s=s_{1b}=\ell/4$ and $\Bigl(\bfr(s_c)+(d/2)\bfn(s_c)-\bfr(0)\Bigr)\cdot \bfe_x=0$. The last condition imposes contacts between the two threads in $O'$. Note that the threads actually touch in two points (Fig.\ref{fig:cross}.c), but the distance between the centerlines in $O'$ is typically $1.03~d\simeq d$, and the distance between contact points is $\ll \ell$, validating the simplifying assumption of an unique contact point at $s_c$. There are no explicit solutions to this 2D-elastica problem, and we solved it numerically. For each arbitrary value of $s_c$ we obtain $f_c$, $\bfr(s_c)$, $\bfn(s_c)$, and braid inclination $\tan(\theta):=-(dy/dx)(s_c)$. Using symmetries of the cell, we obtain:
\begin{subequations}  
\begin{align}
\mathcal{L}_x(s_c)&=\Bigl(4\bigl[\bfr(s_c)-\bfr(\ell/4)\bigr]+2d~ \bfn(s_c)\Bigr)\cdot \bfe_x&\label{eq:lx}\\
    \mathcal{L}_y(s_c)&=\Bigl(2\bigl[\bfr(s_c)-\bfr(0)\bigr]+d~\bfn(s_c)\Bigr)\cdot \bfe_y&\label{eq:ly} 
\end{align}
\end{subequations}  
$(\mathcal{L}_x(s_c),\mathcal{L}_x(s_c))$ is the parametric curve in the plane $(\ell_x,\ell_y)$ shown on Fig.\ref{fig:compa}. All the rest states stand very close to this parametric curve, from a simple 2D approximation of the full 3D problem.

\begin{figure}[t]
\centering
\includegraphics[width=\columnwidth]{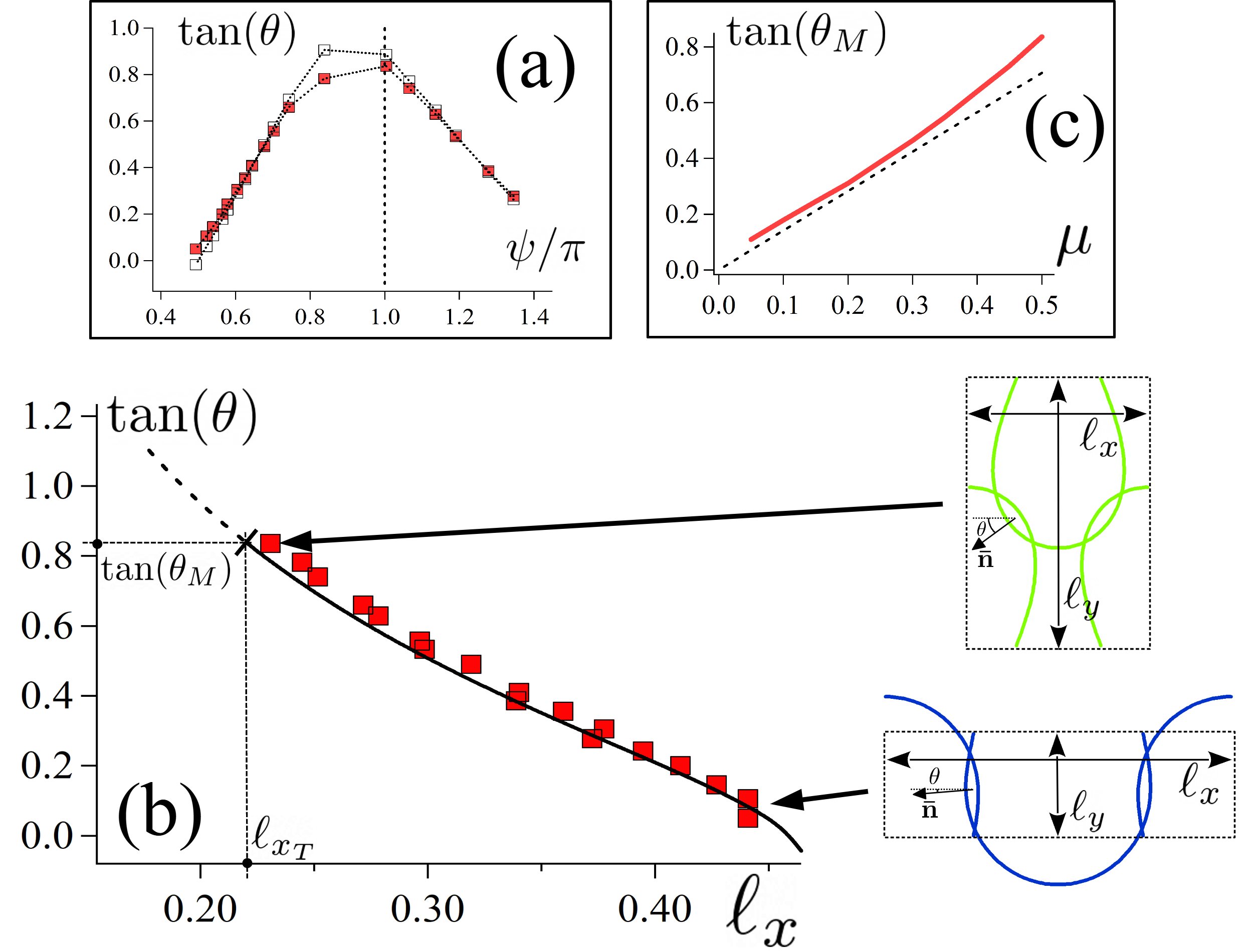}
\caption{(a) For rest states with $\mu=0.5$, braid inclination $\tan(\theta)$ against tangential forces direction $\psi/\pi$. (Filled square): $\tan(\theta)$ measured from DER. (Open square): right term of \eqref{eq:tantheta} with $\varphi$ from DER. (Dashed line): approximated position of the maximum at $\psi=\pi$. (b) Braid inclination $\tan(\theta)$ vs cell dimension $\ell_x$. (Filled squares): DER rest states obtained with $\mu=0.5$. (Line): $\tan(\theta)$ vs $\mathcal{L}_x$ of the 2D-Elastica model. (Black dash): theoretical terminal point. (Insets): threads centerlines for rest states with $\tan(\theta)\simeq 0$ and $\tan(\theta) \simeq \tan(\theta_M)$. Dashed rectangles are the meshes of sizes $\ell_x \times \ell_y$. (c) (Red line): $\tan(\theta_M)$ versus $\mu$ from DER. (Dash line): Eq.\eqref{eq:tantheta.2} with $\varphi = 3 \pi/4$. }
\label{fig:eq6}
\end{figure}

All configurations of the parametric curve verify $\bff_c\cdot \bfe_y=0$. However, the tangential and normal components of contact forces are linked by Coulomb's friction law. This imposes a maximal braid inclination $\tan(\theta_M)$, that sets the terminal point $T(\mu)$. The position of this point is predicted by considering the details of the contact force distribution $\bfp(s)$. 

The braid comprises two twisted fibers in a geometry similar to the one occurring in knots where the threads are twisted of $n$ turns. The cases $n=1$ ($3_1$ knot)~\cite{audoly_2007,clauvelin_2009}, $n=2$ ($5_1$ knot)~\cite{clauvelin_2009} and $n\gg 1$~\cite{jawed_2015} have been considered previously. Threads in the braid are approximately a helix contacting in a few points. It seems that our braid with $n=1/2$ has not been previously considered, but it exhibits a very similar behavior. The threads have nearly helical shapes twisted around a common line of direction $\bftbar=\bigl[\bft_1(s_1)+\bft_2(s_2)\bigr]/\Vert \bft_1(s_1)+\bft_2(s_2) \Vert$, where $\bft_1$ and $\bft_2$ are threads (1) and (2) centerlines' tangent vector. The figure~\ref{fig:cross}.c represents the relative position $\bfr_2(s_2)-\bfr_1(s_1)$ of the threads, where $\bfr_1(s_1)$ and $\bfr_2(s_2)$ are the projections of the centerlines on the plane perpendicular to $\bftbar$. With the fibers contacting in two points $\alpha$ and $\beta$: 
\begin{equation}
    \bff_c=\int \bfp(s) ds = \sum_{\xi=\alpha,\beta} \bigl[f_\xi^{(n)}\bfN_\xi+f_\xi^{(t)}\bfT_\xi\bigr]\label{eq:fc2}
\end{equation}
with $\bfN_\xi$ the contact point's normal vector, $\bfT_\xi$ the orientation of the tangential contact forces (Fig.\ref{fig:cross}(d)), and $f_\xi^{(n)}$ and $f_\xi^{(t)}$ the normal and tangential forces. For all equilibrium configurations, we found $f_\alpha^{(n)}=f_\beta^{(n)}:=f^{(n)}$ and $(\bfN_\alpha +\bfN_\beta) \cdot \bfe_z=0$. Noting $\varphi_\xi$ the angle between $\bfN_\xi$ and $z=0$ (Fig.\ref{fig:cross}.c), we have $\varphi_\alpha=-\varphi_\beta:=\varphi$, with $\varphi\simeq 3 \pi/4$ for each relaxed configuration.

When the meshes are steadily deformed, the threads in the braid slide. The friction is then fully mobilized, and $f_\xi^{(t)}=\mu  f_\xi^{(n)}$. Since $\bfN_\xi\cdot\bft_i(s_\xi)=0$ for the two threads $i=1,2$, $\bfN_\xi$ is perpendicular to the plane containing $\bft_1(s_\xi)$ and  $\bft_2(s_\xi)$. $\bfT_\xi$ belongs to this plane, and we may write:
\begin{equation}
    \bfT_\xi = \cos(\psi_\xi)~\bftbar +  \sin(\psi_\xi)~ \bigl[\bftbar \times \bfN_\xi\bigr] \label{eq:txi}
\end{equation}
where $\psi_\xi$ is the angle between $\bfT_\xi$ and $\bftbar$ (Fig.\ref{fig:cross}.d). Using Eq.\eqref{eq:fc2} and Eq.\eqref{eq:txi}, we may calculate $\bff_c$. With $\varphi_\alpha=-\varphi_\beta$, the equilibrium conditions $\bff_c \cdot \bfe_z=0$ leads to $\psi_\alpha = -\psi_\beta:= \psi$, and finally:
\begin{equation}
    \bff_c = 2 f^{(n)} \times\Bigl(
    \mu \cos(\psi)\bftbar+
\bigl[\cos(\varphi)+\mu\sin(\varphi)\sin(\psi)\bigr]\bfnbar \Bigr)\label{eq:fc3}
\end{equation}

With $\bftbar\cdot\bfe_y=\cos(\theta)$, the equilibrium condition for a $\sigma^{\mathrm{ext}}=0$ configuration, $\bff_c\cdot\bfe_y=0$, writes:
\begin{equation}
    \tan(\theta) = \frac{\mu ~ cos(\psi)}  {\cos(\varphi) +\mu \sin(\varphi) \sin(\psi)}\label{eq:tantheta}
\end{equation}

For a given $\mu$, and because $\varphi$ is roughly constant, variation of $\tan(\theta)$ are driven by variation of the tangential forces direction $\psi$. Fig.\ref{fig:eq6}(a) shows $\tan(\theta)$ versus $\psi$ for the relaxed states obtained by DER simulations and confirms that \eqref{eq:tantheta} holds. The tangential forces at the two contact points $\alpha$ and $\beta$ rotate on either side of the braid axis $\bftbar$. The effect of those rotations is to vary the amplitude of the total friction force along the $\bftbar$ axis, even if individual friction forces always evolve at the Coulomb threshold. 

The inclination of the braid is a fundamental quantity that controls the cell size $(\ell_x, \ell_y)$. Fig.\ref{fig:eq6}(b) shows the relation between $\tan(\theta)$ and $\ell_x$ (similar relation may be obtained for $\ell_y$) for the 2D-elastica model and DER data. The way increasing the inclination $\tan(\theta)$ shrinks $\ell_x$ (Fig.5(b), insets) is well captured by our 2D-elastica model.

The existence and position of the terminal point $T= (\ell_{x_T},\ell_{y_T})$ are predicted by Eq.(6): $\tan(\theta)$ is bounded by some maximal value $\tan(\theta_M)$, the relation between $\tan(\theta)$ and $\ell_x$ (respectively $\ell_y$) imposes a minimum (resp. maximum) value $\ell_{x_T}$ (resp. $\ell_{y_T}$). With $\varphi$ constant, $\tan(\theta)$ is maximal for a rotation $\psi=\pi+\arcsin(\mu\tan(\varphi))\simeq \pi$ for $\mu\ll 1$ and $\varphi \simeq 3 \pi/4$ as shown in Fig.\ref{fig:eq6}.a. This maximal value of $\theta$ is obtained when the frictional forces are roughly aligned with the common tangent of the centerline. Approximating $\psi=\pi$ in Eq.\eqref{eq:tantheta}, we finally obtain for small $\mu$:
\begin{equation}
\tan\theta\le \tan(\theta_M):=\frac{\mu}{\vert\cos(\varphi)\vert}\label{eq:tantheta.2}
\end{equation}
where $\theta_M$ is the maximal value of $\theta$. The simplified prediction of Eq.\eqref{eq:tantheta.2} is verified by confronting, for different values of $\mu$, the predicted values of $\tan(\theta_M)$ with a constant $\varphi=3 \pi/4$ and the ones measured from DER (Fig.\ref{fig:eq6}.c). 

To summarize, in the configuration space $(\ell_x,\ell_y)$, any rest knitted Jersey lies on a curve given by Eq.(2) which does not depend on $\mu$. However, for a given value of $\mu$, only a subset of this curve is possible because of Coulomb's friction law. This subset increases with friction and ends at a point $T(\mu)$ for which $\theta=\theta_M(\mu)$ (fixed by Eq.(7)).\\

Our study rationalizes the $\sigma^{\mathrm{ext}}=0$ rest state of a periodic yarn assembly. They are not unique but form a continuous subset, bounded by a terminal point $T$, in the space of possible periodic configurations $(\ell_x,\ell_y)$ of a knitted fabric.
Those findings have many implications.  
The configuration corresponding to $T$ would be the reproducible shape of a $\sigma^{\mathrm{ext}}=0$ knitted fabric, accessible through successive stretching cycles along $y$, even if metastability makes other rest shapes possible. 
 The existence of a continuum of relaxed states has important consequences for the macroscopic mechanical properties. The restoring forces are, therefore, weak over a wide range of the configuration space. Knitted fabrics are soft objects for deformations that remain in this zone but are relatively rigid when we move away from it. Finally, variations in aspect ratios mean that the area per stitch $\ell_x\times\ell_y$ can be varied at zero external force. A flat knitted fabric can thus be stretched to cover a surface with non-zero Gaussian curvature without any external forces being applied. 
 This study can also serve as a ground basis for exploring further the mechanics of knitted fabrics or, more generally, of periodic three-dimensional structures ~\cite{greer_2023}. The numerical and theoretical models can be adapted to different knitting topology~\cite{singal_programming_2024}, but also for fibers closer to applications by modifying the elastic properties of the rod or smaller aspect ratio $\ell/d$. The methods introduced here can also be adapted to explore the role of friction in the force vs. strain responses of textiles, including hysteresis~\cite{poincloux.2018b} and slip-induced fluctuations~\cite{poincloux.2018}. 

\begin{acknowledgments}
A.S. thanks D. Le Tourneau and P. Metz for building the biaxial tensile machine. J.C. acknowledges CNRS-Physique for hosting in delegation. S.P. acknowledges financial support from the Japanese Society for the Promotion of Science as a JSPS International Research Fellow. This work has been supported by Agence Nationale de la Recherche Grant ANR-23-CE30-0015.
\end{acknowledgments}

\bibliography{references}
\bibliographystyle{unsrt}
\end{document}